\documentclass[12pt]{article}
\usepackage{epsfig}
\begin{document}
\begin{titlepage}
\begin{center}

\hfill{SNUTP/06-007}

\vspace{2cm}

{\Large \bf Instanton interpolating current for
$\sigma$--tetraquark} \\
\vspace{0.50cm}
\renewcommand{\thefootnote}{\fnsymbol{footnote}}
Hee-Jung Lee$^{a,b}$ \footnote{hjlee@www.apctp.org}, N.I.
Kochelev$^{b,c,d}$\footnote{kochelev@theor.jinr.ru}

\vspace{0.50cm}

{(a) {\it Department of Science Education, Chungbuk National
University,\\
Cheongju Chungbuk 361-763, South Korea}} \vskip 1ex {(b) {\it Asia
Pacific Center for Theoretical Physics, POSTECH,\\
Pohang 790-784, South Korea}}\vskip 1ex
{(c) {\it School of Physics and  Astronomy and Center for Theoretical Physics,\\
Seoul National University, Seoul 151-747, South Korea}} \vskip 1ex
{(d) {\it Bogoliubov Laboratory of Theoretical Physics,\\
Joint Institute for Nuclear Research, Dubna, Moscow region, 141980 Russia}}

\end{center}
\vskip 0.5cm

\centerline{\bf Abstract} We perform a QCD sum rule analysis for
the light scalar meson $\sigma$ ($f_0(600)$) with a tetraquark
current related to the instanton picture for QCD vacuum.  We
demonstrate that instanton current, including equal weights of
scalar and pseudoscalar diquark-antidiquarks, leads to a strong
cancelation between the contributions of high dimension operators
in the operator product expansion (OPE). Furthermore, in the case
of this current direct instanton contributions  do not spoil the
sum rules. Our calculation, obtained from the OPE up to dimension
10 operators, gives the mass of $\sigma$--meson around 780MeV.

 \vskip 0.3cm \leftline{Pacs: 11.55.Hx, 12.38.Aw, 12.38.Lg,
14.40-n} \leftline{Keywords: QCD sum rule, scalar meson,
tetraquark, instanton, OPE}

\vspace{1cm}

\end{titlepage}

\setcounter{footnote}{0}
\section{Introduction}
Nowadays there is a lot of controversy in the interpretation of
the scalar mesons with the masses below 1 GeV~\cite{Amsler04}. In
the constituent quark models, they are expected to have the quark
content of $q\bar{q}$ as the normal members of flavor $SU(3)_f$
nonet with one unit of orbital excitation for positive parity.
However, from the fact that the orbital excitation gains energy
about 0.5 GeV, it is difficult to explain their light masses as
well as their mass spectrum (so called inverted mass
spectrum)~\cite{Jaffe}. Additionally, the two candidates to the
members of nonet, the isovector $a_0$(980) and isoscalar
$f_0$(980), have a very peculiar properties. Indeed, their masses
are degenerated and they have strong coupling to $K\bar{K}$
channel in strong contradiction with expectation of simple $q\bar
q$ picture of the mesons. This puzzle stimulated alternative
interpretations of these mesons as various types of tetraquark
states \cite{Jaffe2}, e.g., meson--meson  molecule
states~\cite{KK}, diquark--antidiquark bound
states~\cite{Jaffe,Alford,Maiani04,Brito05,chinese05}, and as some
hybrid states of mixture of $q\bar{q}$ and
meson--meson~\cite{Anisovich05} and of diquark--antidiquark and
meson--meson~\cite{Close02}.

In this paper, we will consider the properties of the lightest
scalar meson state $\sigma$ ($f_0(600)$) as a diquark--antidiquark
bound state within the QCD sum rule~\cite{QCDSR} with the special
choice of the tetraquark current. We should mention that the QCD
sum rules (SR) for the light scalar mesons  were already
considered separately with the interpolating currents of the
scalar diquark--antidiquark and  of the pseudoscalar
diquark--antidiquark. In these calculations only contributions  to
OPE up to the operators of dimension 6  have been
considered~\cite{Brito05,chinese05}. However, it was recently
shown in \cite{Lee06,Lee05,Ioffe04} that SR for multiquark systems
might receive a large contributions from the operators of higher
dimensions which can lead  to strong instability  of the obtained
results for some types of interpolating currents. In particularly,
it has been demonstrated that in the case of the scalar
diquark--scalar antidiquark interpolating current for light
tetraquarks, the contribution of the operators of dimensions 8
violates the requirement of positivity of left-hand side (LHS) and
leads to disappearance of the bound state tetraquark signal
\cite{Lee05}.

In this Letter  we  suggest  to use special type of interpolating
tetraquark current for lightest scalar meson  $\sigma$
 which  leads  simultaneously to the
cancelation of high dimensional condensate contributions to the
OPE and some dangerous instanton contribution to SR. The
$\sigma$--state has the vacuum quantum numbers and should couple
to the QCD vacuum very strongly. Instantons, topological
fluctuations of gluon fields, play very important role in
structure of QCD vacuum \cite{shuryak98} and in  spectroscopy of
the multiquark  hadrons \cite{dorkoch}, \cite{schafer},
\cite{LKV}.
 Therefore, our basic idea  is to use the
color--spin--flavor structure of the four--quark interaction
induced by instantons~\cite{Rapp98} to fix the possible ``good"
interpolating current for $\sigma$--state. This specific
interaction gives strong correlations between scalar diquark of
$\bar{\bf 3}_c$ and scalar antidiquark of ${\bf 3}_c$, between
pseudoscalar diquark of $\bar{\bf 3}_c$ and pseudoscalar
antidiquark of ${\bf 3}_c$, and between tensor diquark of
$\bar{\bf 6}_c$ and tensor antidiquark of ${\bf 6}_c$, where the
subscript $c$ means color. With this inspection, we propose that
the interpolating current for $\sigma$(600) consists of the above
three types of diquark--antidiquark combinations. Our additional
argument in favor of instanton current is based on the phenomena
of the cancelation of high dimension operator contributions
expected for the case of the OPE in the self--dual vacuum fields
\cite{smilga}. Instanton field is a self--dual field and  this
self--duality property manifests directly in the
color--spin--flavor structure of the four--quarks instanton
interaction. Therefore, we can expect a similar cancelation for
the case of tetraquark current obtained from the quark--quark
instanton induced interaction.

We  construct the QCD SR with the OPE up to operators of dimension
10 and show that  for the case of instanton current with equal
weights for the scalar diquark--antidiquark and the pseudoscalar
diquark--antidiquark, the cancelation takes place separately  for
the  high dimension operator contributions and for some dangerous
instanton contributions to the SR. Our results for the left--hand
side (LHS) of the SR are very stable. For the phenomenological,
right--hand side (RHS) of the SR, we apply the two resonance
approximation for the spectral representation, which allows to
avoid the well known problem of strong dependence of the results
for multiquark systems on the value of
threshold~\cite{Lee06,Narison04}.

This paper is organized as follows.  In Sec.~II, using the
instanton induced quark-quark interaction, we fix the general
structure of interpolating current for $\sigma$--tetraquark. In
Sec.~III we construct the standard QCD sum rule for $\sigma$ based
on the OPE. In Sec.~IV the direct instanton effects to the sum
rule are considered. We present the numerical results in Sec.~V
and discuss them in the Conclusion.

\section{The instanton current for $\sigma$-tetraquark}

The famous 't Hooft instanton induced interaction between light
quarks~\cite{tHooft76} for case of $N_f=2$ can be written in
following form~\cite{Rapp98}
\begin{eqnarray}
{\cal L}=\frac{G}{4(N_c^2-1)}\bigg[
\frac{2N_c-1}{2N_c}\bigg((\bar{\psi}\tau^-_\mu\psi)^2
+(\bar{\psi}\gamma_5\tau^-_\mu\psi)^2\bigg)
+\frac{1}{4N_c}(\bar{\psi}\sigma_{\rho\sigma}\tau^-_\mu\psi)^2\bigg]\
\label{lag}
\end{eqnarray}
where $\psi$ is two flavors spinor, $N_c$ is the number of colors
and $\tau^-_{\mu}=(\vec{\tau},i).$ This Lagrangian can be
transformed to the Lagrangian for the interactions between diquark
and antidiquark by a Fierz transformations in the spin, flavor and
color spaces :
\begin{eqnarray}
{\cal L}
&=&-\frac{G}{8N_c(N_c-1)}\bigg[
({\psi}^TC\gamma^5\tau_2\lambda^A\psi)
(\bar{\psi}\tau_2\lambda^A\gamma^5C\bar{\psi}^T)
+({\psi}^T\tau_2\lambda^AC\psi)(\bar{\psi}\tau_2\lambda^A C\bar{\psi}^T)\bigg]
\nonumber\\
&&+\frac{G}{16N_c(N_c+1)}({\psi}^T\tau_2\lambda^SC\sigma^{\rho\sigma}\psi)
(\bar{\psi}\tau_2\lambda^S\sigma_{\rho\sigma}C\bar{\psi}^T)\ ,
\end{eqnarray}
where $\lambda^{A,S}$ are the antisymmetric and symmetric color
generators normalized as ${\rm
Tr}(\lambda^a\lambda^b)=2\delta^{ab}$, respectively. The first two
terms correspond to the scalar and the pseudoscalar diquarks in
antisymmetric $\bar{\bf 3}_c$ representation and the last one to the
tensor diquark in symmetric $\bar{\bf 6}_c$ color state.

By introducing the spin matrices,
\begin{equation}
\Gamma_S=C\gamma^5, \ \
\Gamma_{PS}=C,\ \ \Gamma_{T,\rho\sigma}=C\sigma_{\rho\sigma}\ ,
\end{equation}
we can rewrite the Lagrangian in terms of flavors
\begin{eqnarray}
{\cal L}
&=&\frac{G}{2N_c(N_c-1)}\ \epsilon_{abc}\epsilon_{ade}\bigg[
(u^T_b\Gamma_{S}d_{c})(\bar{u}_d\overline\Gamma_{S}\bar{d}^T_{e})
-(u^T_b\Gamma_{PS}d_{c})(\bar{u}_d\overline\Gamma_{PS}\bar{d}^T_{e})\bigg]
\nonumber\\
&&+\frac{G}{4N_c(N_c+1)}(u^T_a\Gamma_{T,\rho\sigma}d_{a'})
\bigg((\bar{u}_a\overline{\Gamma}_T^{\rho\sigma}\bar{d}^T_{a'})
+(\bar{u}_{a'}\overline{\Gamma}_T^{\rho\sigma}\bar{d}^T_{a})\bigg)\ ,
\label{Ldi}
\end{eqnarray}
where $\overline{\Gamma}_i=\gamma^0\Gamma_i^\dagger\gamma^0$ and the properties
\begin{equation}
\overline{\Gamma}_S=-C\gamma^5,\ \
\overline{\Gamma}_{PS}=C,\ \
\overline{\Gamma}_{T,\rho\sigma}=-\sigma_{\rho\sigma}C,
\end{equation}
have been used. One can see that only a very restricted set of
diquarks can strongly interact with instanton field, namely scalar,
pseudoscalar and tensor diquarks.

Therefore we suggest the following interpolating current for the
$\sigma$--meson :
\begin{equation}
J^\sigma=\alpha J^\sigma_S+\beta J^\sigma_{PS}+\gamma J^\sigma_{T}
\label{current}
\end{equation}
where each current is defined by
\begin{eqnarray}
J^\sigma_S&=&\epsilon_{abc}\epsilon_{ade}
(u^T_b\Gamma_{S}d_{c})(\bar{u}_d\overline\Gamma_{S}\bar{d}^T_{e})\ ,
\nonumber\\
J^\sigma_{PS}&=&\epsilon_{abc}\epsilon_{ade}
(u^T_b\Gamma_{PS}d_{c})(\bar{u}_d\overline\Gamma_{PS}\bar{d}^T_{e})\ ,
\nonumber\\
J^\sigma_T&=&(u^T_a\Gamma_{T,\rho\sigma}d_{a'})
(\bar{u}_a\overline{\Gamma}_T^{\rho\sigma}\bar{d}^T_{a'}
+\bar{u}_{a'}\overline{\Gamma}_T^{\rho\sigma}\bar{d}^T_{a}),
\end{eqnarray}
where $\alpha, \beta$ and $\gamma$ are some constants. From the
Lagrangian Eq.~(\ref{Ldi}), for $N_c=3$, it is expected that the
ratio of the coefficients
\begin{equation}
\alpha : \beta : \gamma=1:-1:\frac{1}{4} \label{ratio}
\end{equation}
may provide some specific properties of OPE expansion for the $\sigma$ correlator
and may finally lead to the most stable QCD sum rule.

\section{The OPE contribution to $\sigma$-meson correlator}

\begin{figure}[h]
\centerline{\epsfig{file=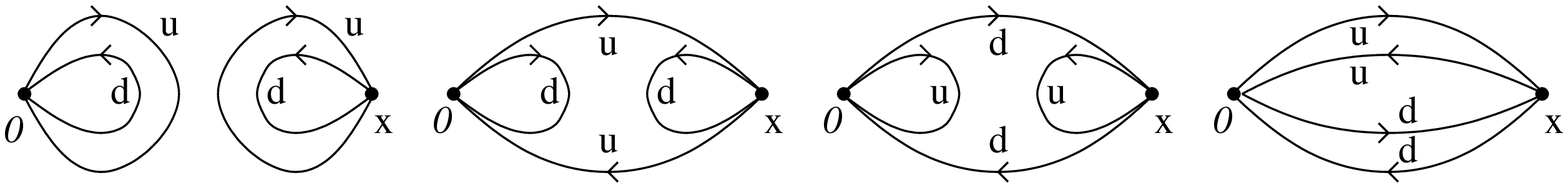,width=12cm,angle=0}}
\caption{Diagrammatic representation of the various correlators
Eq~(\ref{corr1}). Each quark line means the full quark
propagator.} \label{corr}
\end{figure}

The $\sigma$--correlator for the case of current Eq.~(\ref{current})
is decomposed into nine parts
\begin{eqnarray}
\Pi^\sigma&=&i\int d^4x e^{iq\cdot x}
\langle0|TJ^\sigma(x)J^{\sigma\dagger}(0)|0\rangle
\nonumber\\
&=&\alpha^2\Pi^{S,S}+\beta^2\Pi^{PS,PS}+\gamma^2\Pi^{T,T}
\nonumber\\
&&+\alpha\beta(\Pi^{S,PS}+\Pi^{PS,S})+\alpha\gamma(\Pi^{S,T}+\Pi^{T,S})
+\beta\gamma(\Pi^{PS,T}+\Pi^{T,PS})\ . \label{corr1}
\end{eqnarray}
$\Pi^{A,B}$ means the correlator between $A$--type current and
$B$--type current. Full set of  the diagrams for the
$\sigma$--correlator is presented in Fig.~\ref{corr}. It is
evident that we should consider only the last diagram contribution
to SR since only that diagram  is relevant in the description of
$\sigma$ as the tetraquark state. The propagator for massless
quarks $q=u, d$ in the fixed point gauge in Fig.~\ref{corr}  up to
order of the $g^2$ in strong coupling constant is the
following~\cite{propagator}
\begin{eqnarray}
S^q_{ab}(x)&=&-i\langle 0|Tq_a(x)\bar{q}_b(0)|0\rangle
\nonumber\\
&=&\delta_{ab}\bigg(\frac{\hat{x}}{2\pi^2 x^4}
+i\frac{\langle \bar{q}q\rangle}{12}
-\frac{x^2}{192}\langle g\bar{q}\sigma\cdot G q\rangle
+i\frac{x^4}{2^{9}\cdot3^3}\langle \bar{q}q\rangle \langle g^2G^2\rangle\bigg)
\nonumber\\
&&-i\frac{g}{32\pi^2}G^{\mu\nu}_{ab}\frac{1}{x^2}
(\hat{x}\sigma_{\mu\nu}+\sigma_{\mu\nu}\hat{x}),
\end{eqnarray}
where $a,b$ are the color indices.

\begin{figure}[h]
\centerline{\epsfig{file=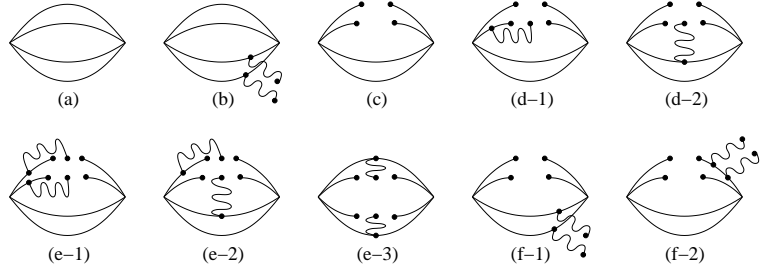,width=10cm,angle=0}}
\caption{Diagrammatic representation of Eq.~(\ref{Im}).}
\label{SR}
\end{figure}

The OPE, up to the operators of dimension 10, yields the imaginary
part of the $\sigma$--correlator
\begin{eqnarray}
\frac{1}{\pi}\ {\rm Im}\
\Pi^\sigma_{OPE}(q^2)&=&(\alpha^2+\beta^2+48\gamma^2)\frac{(q^2)^4}{2^{12}\cdot5\cdot3\pi^6}\
\bigg|_{(a)}
\nonumber\\
&&+(\alpha^2+\beta^2+88\gamma^2+12\alpha\gamma
-12\beta\gamma)\frac{\langle g^2
G^2\rangle}{2^{11}\cdot3\pi^6}(q^2)^2\bigg|_{(b)}
\nonumber\\
&&+(\alpha^2-\beta^2)\frac{\langle\bar{q}q\rangle^2}{12\pi^2}\
q^2\bigg|_{(c)}-(\alpha^2-\beta^2)\frac{\langle\bar{q}q\rangle
\langle ig\bar{q}\sigma\cdot Gq\rangle}{12\pi^2}\
\bigg|_{(d)}\nonumber\\
&&+(\alpha^2-\beta^2)\frac{59(\langle ig \bar{q}\sigma\cdot
Gq\rangle)^2}{2^{9}\cdot3^2\pi^2}\
\delta(q^2)\bigg|_{(e)}
\nonumber\\
&&+(\alpha^2-\beta^2)\frac{7\langle g^2
G^2\rangle\langle\bar{q}q\rangle^2}{2^5\cdot3^3\pi^2}\ \delta(q^2)
\bigg|_{(f)}\ \ , \label{Im}
\end{eqnarray}
where each term corresponds to each diagram shown in Fig~\ref{SR}
and the factorization hypothesis for high dimension operators has
been used.

\section{The direct instanton contribution}

\begin{figure}[h]
\centerline{\epsfig{file=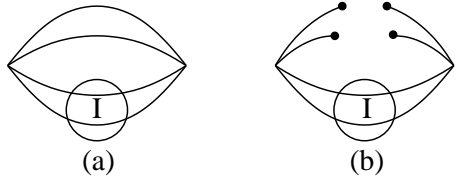,width=6cm,angle=0}}
\caption{Direct instanton contributions to the correlator.}
\label{inst}
\end{figure}

In addition to contributions of power type from the OPE expansion to
the QCD SR, there are exponential contributions coming from direct
instantons contributions as shown in Fig.~\ref{inst}. Their
contributions can be calculated by using the following formula in
Euclidean space for the quark propagator on the instanton background
in the regular gauge
\begin{equation}
S^{q,inst}_{ab}(x,y)=A_q(x,y)\gamma_\mu\gamma_\nu (1+\gamma_5)
(U\tau^-_\mu\tau^+_\nu U^\dagger)_{ab},
\end{equation}
where $$ A_q(x,y)=-i\frac{\rho^2}{16\pi^2
m_q^{*}}\phi(x-z_0)\phi(y-z_0) $$
and
$$\phi(x-z_0)=\frac{1}{[(x-z_0)^2+\rho^2]^{3/2}}.$$
Here $\rho$ stands for the instanton size, $z_0 $ for the center
of the instanton. $U$ represents the color orientation matrix of
the instanton in $SU(3)_c$ and $\tau_{\mu,\nu}^{+,-}$ are
$SU(2)_c$ matrices. The effective mass of quark on the instanton vacuum is
$m_q^*=m_q-2\pi^2\rho_c^2\langle\bar q q\rangle/3$ with current quark mass $m_q$.
At the final stage, we multiply the result by a factor of
two to take into account the anti--instanton contribution and
integrate over the color orientation and the instanton size.

With the definition $Q^2=-q^2$, the direct instanton contribution to
the $\sigma$--correlator from the two diagrams in Fig.~\ref{inst}
is given by
\begin{eqnarray}
\Pi^\sigma_{I+\bar{I}}(Q)&=&(\alpha^2-\beta^2)\frac{32n_{eff}\rho_c^4}
{\pi^8m_q^{*2}}f_6(Q)
\nonumber\\
&&+[19(\alpha^2+\beta^2)-6\alpha\beta+912\gamma^2+72\alpha\gamma+72\beta\gamma]
\frac{n_{eff}\rho_c^4\langle\bar{q}q\rangle^2}{18\pi^4m_q^{*2}}
f_0(Q),
\label{instSR}
\end{eqnarray}
where Shuryak's instanton liquid model for QCD vacuum with density
$n(\rho)=n_{eff}\delta(\rho-\rho_c)$~\cite{shuryak98} has been used
and $\bar{I}$ means the contribution from anti--instanton.
$f_6(Q),\ f_0(Q)$ are the functions defined by
\begin{eqnarray}
f_6(Q)&=&\int d^4z_0 \int d^4x \frac{e^{iq\cdot x}}
{x^6[z_0^2+\rho_c^2]^3[(x-z_0)^2+\rho_c^2]^3}\ ,
\nonumber\\
f_0(Q)&=&\int d^4z_0 \int d^4x \frac{e^{iq\cdot x}}
{[z_0^2+\rho_c^2]^3[(x-z_0)^2+\rho_c^2]^3}\ . \label{func}
\end{eqnarray}
Here $\rho_c$ is the average instanton size.  Moreover,  let us
also note that direct instanton contribution is possible only for
the different quark flavors. Therefore, in case of tetraquark
$\sigma$ meson with pure $u$-- and $d$-- quark content, there is
no direct instanton contribution in any subsystem of three quarks.
On the other hand, three--body instanton contribution in $\bar ud
s, u\bar ds, ud\bar s$ subsystems might be quite important in the
case of $f_0(980)$ and $a_0(980)$ tetraquarks. Furthermore, such
interaction can also lead to sizeable mixing of $ud\bar u\bar d$
$\sigma$--meson with convenient $s\bar s$ two--quark state.

\section{Numerical analysis of QCD sum rule for instanton current}

In order to avoid strong dependence of the
 multiquark mass on the value of threshold~\cite{Narison04}, we
apply the two resonances approximation to the spectral
representation of the correlator as
\begin{eqnarray}
{\rm Im}\ \Pi^\sigma(s^2)&=&\pi\sum_n\delta(s^2-m_n^2)\langle 0|J^\sigma|n\rangle
\langle n|J^{\sigma\dagger}|0\rangle
\nonumber\\
&=&2\pi f_{1}^2m_{1}^8\delta(s^2-m_1^2)
+2\pi f_{2}^2m_{2}^8\delta(s^2-m_2^2)+\theta(s^2-s_0^2){\rm Im}\ \Pi^{OPE}(s^2)
\label{stSR}
\end{eqnarray}
with the convention
\begin{equation}
\langle0|J^{\sigma}|S_i\rangle=\sqrt{2}f_im_i^4\ .
\end{equation}
In this case the QCD sum rule is the following
\begin{eqnarray}
&&(\alpha^2+\beta^2+48\gamma^2)\frac{M^{10}E_4}{2^9\cdot 5\pi^6}
+(\alpha^2+\beta^2+88\gamma^2+12\alpha\gamma-12\beta\gamma)  \frac{\langle g^2G^2\rangle M^6E_2}{2^{10}\cdot
3\pi^6}
\nonumber\\
&&+(\alpha^2-\beta^2)\frac{\langle\bar{q}q\rangle^2}{12\pi^2}M^4E_1
-(\alpha^2-\beta^2)\frac{\langle\bar{q}q\rangle \langle ig\bar{q}\sigma\cdot
Gq\rangle}{12\pi^2}M^2E_0
\nonumber\\
&&+ (\alpha^2-\beta^2)\frac{59(\langle ig\bar{q}\sigma\cdot
Gq\rangle)^2}{2^{10}\cdot3^2\pi^2}
+(\alpha^2-\beta^2)\frac{7\langle g^2
G^2\rangle\langle\bar{q}q\rangle^2}{2^6\cdot3^3\pi^2}
+(\alpha^2-\beta^2)\frac{32n_{eff}\rho_c^4}
{\pi^8m_q^{*2}}\hat{B}[f_6(Q)]
\nonumber\\
&&+[19(\alpha^2+\beta^2)-6\alpha\beta+912\gamma^2+72\alpha\gamma+72\beta\gamma]
\frac{n_{eff}\rho_c^4\langle\bar{q}q\rangle^2}{18\pi^4m_q^{*2}}
\hat{B}[f_0(Q)]
\nonumber\\
&=&2f_{1}^2m_{1}^8e^{-m_{1}^2/M^2}
+2f_{2}^2m_{2}^8e^{-m_{2}^2/M^2}
\label{SR2}
\end{eqnarray}
up to operators of dimension 10, where $\hat{B}[f_{0,6}(Q)]$ is
the Borel transform $f_{0,6}(Q)$ function given by
Eq.(\ref{func}). The contribution from the continuum is encoded in
the functions $E_n(M)$ defined by
\begin{eqnarray}
E_n(M)&=&\frac{1}{\Gamma(n+1)M^{2n+2}}\int_0^{s_0^2}ds^2\ e^{-s^2/M^2}(s^2)^n\ ,
\end{eqnarray}
where $s_0$ is the threshold of the continuum and $M$ is the Borel
mass. In the calculation of the Borel transformed $f_6(Q)$ we only
include the contribution from the pole at finite distance $x^2\sim
-\rho^2$, which corresponds to the direct instantons effect (see
discussion in ~\cite{Lee06}) :
\begin{eqnarray}
\hat{B}[f_6(Q)]&=&-\frac{\pi^4M^{12}}{2^{13}}\int_0^1dt\int_0^1
dy\ \frac{e^{-M^2\rho_c^2/(4ty(1-y))}}{y^2(1-y)^2}
\bigg(X^2+5X^3+10X^4
\nonumber\\
&&+10X^5+5X^6+X^7\bigg)\ ,
\nonumber\\
\hat{B}[f_0(Q)]&=&\frac{\pi^4M^6}{16}\ e^{-M^2\rho_c^2/2}
\bigg(K_0(M^2\rho_c^2/2)+K_1(M^2\rho_c^2/2)\bigg)\ ,
\end{eqnarray}
where $X=(1-t)/t$ and $K_n(x)$ is the McDonald function.

For the numerical analysis with the massless quarks $q=u, d$, we
use the following condensates at normalization point $\mu=1$GeV
and the average size of instanton,
\begin{eqnarray}
&&\langle\bar{q} q\rangle=-(0.23\ {\rm GeV})^3,\ \ \langle
ig\bar{q}\sigma\cdot Gq\rangle=0.8\ {\rm GeV}^2\langle\bar{q}
q\rangle,\ \
\nonumber\\
&&\langle g^2G^2\rangle=0.5\ {\rm GeV}^4, \ \ \rho_c=1.6{\rm
GeV}^{-1}\ .
\end{eqnarray}
As we already mentioned, the sum rules constructed with only the
scalar diquark--antidiquark, $\beta=\gamma=0$, and with only
pseudoscalar diquark--antidiquark, $\alpha=\gamma=0$,  are not
stable. More precisely, the sum rule with the scalar
diquark--antidiquark looses its physical meaning because the LHS
of the sum rule has definite negative value~\cite{Lee05}. On the
other hand, the LHS of the sum rule with the pseudoscalar
diquark--antidiquark has a definite positive value but its slope
is negative so that it is impossible to fit a physical mass for
the resonance. We present this situation in Figs. 4 and 5 with the
value $s_0=1.0$~GeV which is the threshold taken usually in the
single resonance approximation for the RHS of SR
~\cite{Brito05,Lee05}. The origin of this behavior of the LHS of
the sum rule lies in large contributions from the higher
dimensional operators and direct instantons.

\begin{figure}[h]
\begin{minipage}[c]{8cm}
\hspace*{0.5cm} \psfig{file=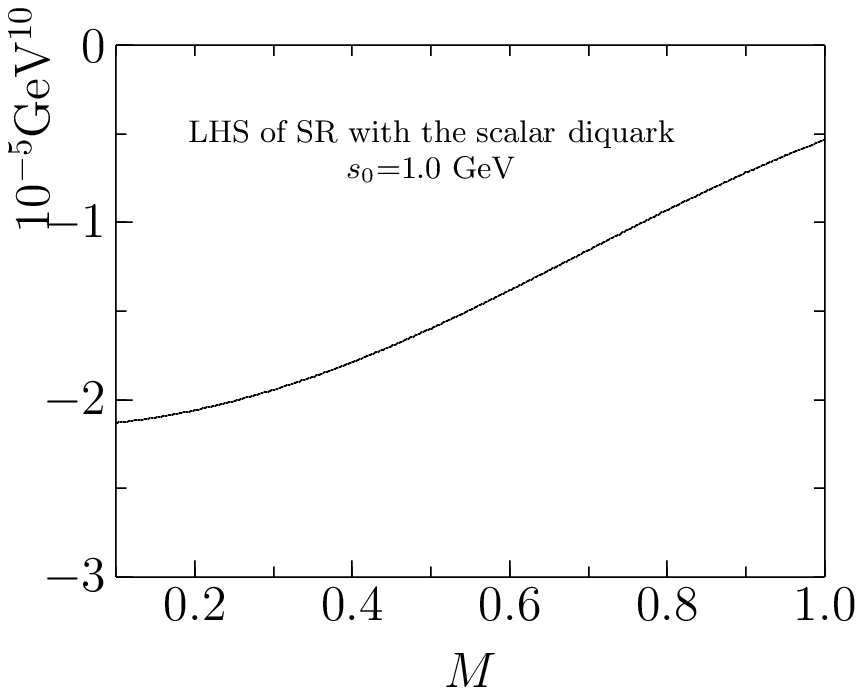,width=6cm,height=5cm}
\caption{The left hand side of the QCD sum rule for $\sigma$ with
the scalar diquark and the scalar antidiquark and $s_0=1.0$GeV.}
\end{minipage}
\hspace*{0.5cm}
\begin{minipage}[c]{8cm}
\hspace*{0.5cm} \psfig{file=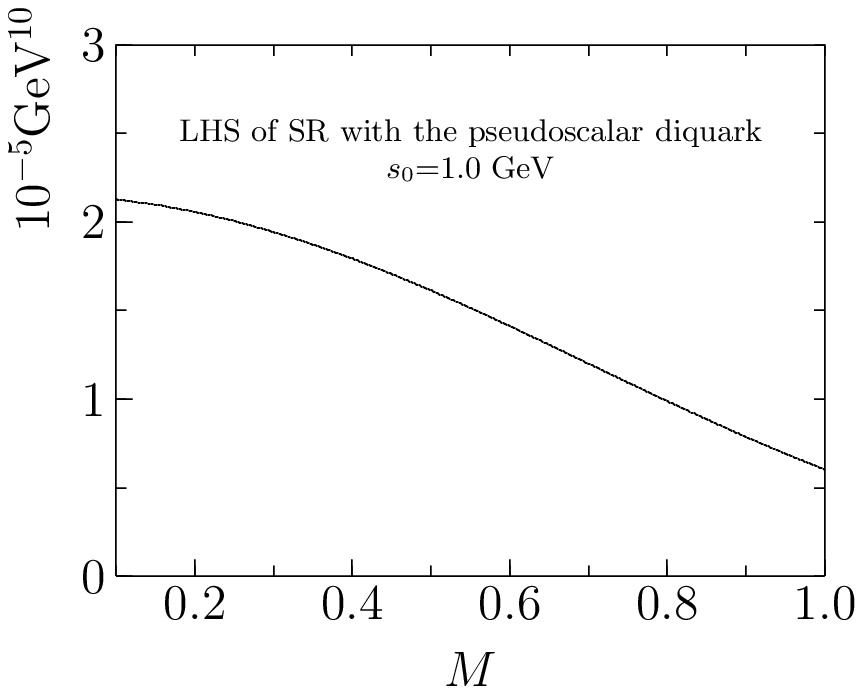,width=6cm,height=5cm}
\caption{The left hand side of the QCD sum rule for $\sigma$ with
the pseudoscalar diquark and the pseudoscalar antidiquark and
$s_0=1.0$GeV.}
\end{minipage}
\end{figure}

\begin{figure}[h]
\centerline{\epsfig{file=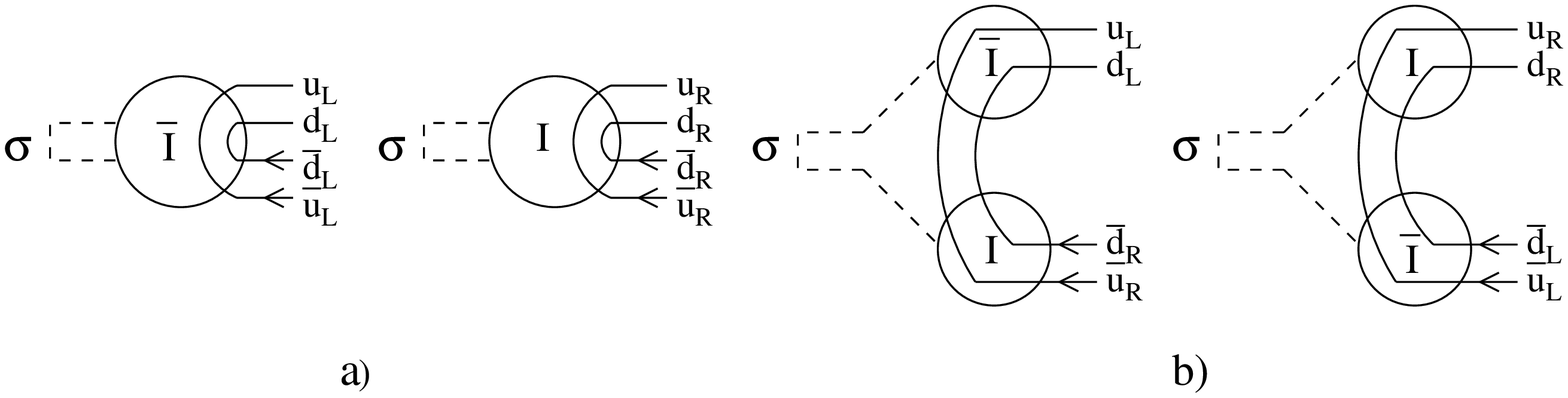,width=15cm,height=4cm,angle=0}}
\caption{ Two suitable currents for $\sigma$-meson a) single
instanton current and b) current induced by
instanton--antiinstanton molecules. The symbol $I$ ($\bar I$)
denote instanton (antiinstanton). } \label{ins}
\end{figure}

Therefore, the pure scalar or pseudoscalar diquark content of
$\sigma$--tetraquark is not favored in the QCD sum rule approach
and we need to find another interpolating current which will be
not so much affected by the higher dimension condensates and the
direct instantons. Let us discuss in detail the origin of the
specific dependence, of the different terms in the LHS of the SR
Eq.(\ref{SR2}), on the parameters of current $\alpha$ and $\beta$.
We would like to emphasize, that the chirality structure of the
current plays an important role in the appearance or disappearance
of some OPE and instanton contributions\footnote{ We should
mention, that from our point of view, the chirality structure of
the multiquark currents might be one of the cornerstones of
multiquark spectroscopy because it provides a strong restriction
on the possible ``good" multiquark currents. Particularly, in
recent papers~\cite{Ioffe04,Oganesian06} it has been shown that
the chirality structure of the pentaquark current is very
important to explain its small width.}. Indeed, the inspection of
our general current Eq.(\ref{current}) shows that the two quarks
(antiquarks) in the diquarks (antidiquarks) in all above currents
should have the same chirality. The appearence of an extra
$\gamma_5$ in the scalar diquark--antidiquark current induces an
additional negative sign between opposite chirality diquarks with
respect to the pseudoscalar diquark--antidiquark current so that
\begin{eqnarray}
\alpha J^\sigma_S+\beta
J^\sigma_{PS}&\sim&-(\alpha-\beta)(u_L^TCd_L\bar{u}_LC\bar{d}_L^T
+u_R^TCd_R\bar{u}_RC\bar{d}_R^T)
\nonumber\\
&&+(\alpha+\beta)(u_R^TCd_R\bar{u}_LC\bar{d}_L^T+u_L^TCd_L\bar{u}_RC\bar{d}_R^T)\
,
\label{chirality}
\end{eqnarray}
where we have dropped the color indices for simplicity. From
Eq.(\ref{current}) it is evident that there are several possible
contributions to the SR with definite chirality flip. Firstly, the
chirality conserved diagonal transitions between four terms in
Eq.(\ref{chirality}) gives the factor $(\alpha^2+\beta^2)$ in
front of chirality conserved OPE terms in the SR. Secondly the
contribution is coming from a non--diagonal transition between the
first and second line in Eq.(\ref{current}). In this case a flip
of chirality of two quarks in the four quark system happens and
the factor $(\alpha^2-\beta^2)$ in front of the high dimension OPE
condensates, and some instanton contributions, appears. There is
also the possibility to have a chirality flip for all quarks
coming from transitions between first (third) and second (fourth)
terms. In OPE expansion this contribution is not vanishing only
for massive quarks. Indeed, one cannot put all quark lines to zero
virtuality in chirality odd condensates due to necessity to have
non--zero momentum transfer through the last diagram in
Fig.\ref{corr}. This momentum transfer with chirality flip in the
corresponding quark line can be produced in the OPE only for a
mass dependent term in the perturbative part of the full quark
propagator. Therefore, for $\sigma$--tetraquark such contribution
is not possible. In addition to the chirality arguments above, the
spin structure of the tensor current
\begin{eqnarray}
J^\sigma_{T}&\sim&-(u_RC\sigma_{\rho\sigma}d_R+u_LC\sigma_{\rho\sigma}d_L)
(\bar{u}_R\sigma^{\rho\sigma}C\bar{d}_R+\bar{u}_L\sigma^{\rho\sigma}C\bar{d}_L)
\end{eqnarray}
restricts further the high dimension OPE contributions from the
tensor current to the sum rule. It is evident from our discussion
above that the cases $\alpha=-\beta$ and $\alpha=\beta$ are very
peculiar cases for the $\sigma$ SR. With these values of $\alpha$
and $\beta$ the contributions from operators with dimension higher
than $d=6$ and some of direct the instanton contributions in
Eq.(\ref{SR2}) vanish. Therefore, for these currents we could
expect more stable QCD SR results. The difference between the two
cases is again in the chirality structure of the corresponding
current. For $\alpha=-\beta$ each term in the first line in
Eq.(\ref{chirality}) has four units of chirality, while for
$\alpha=\beta$ each term in second line in Eq.(\ref{chirality})
has zero chirality. From the point of view of the instanton model
for the QCD vacuum the two cases presented in Fig.\ref{ins}.

\begin{figure}[h]
\centerline{\epsfig{file=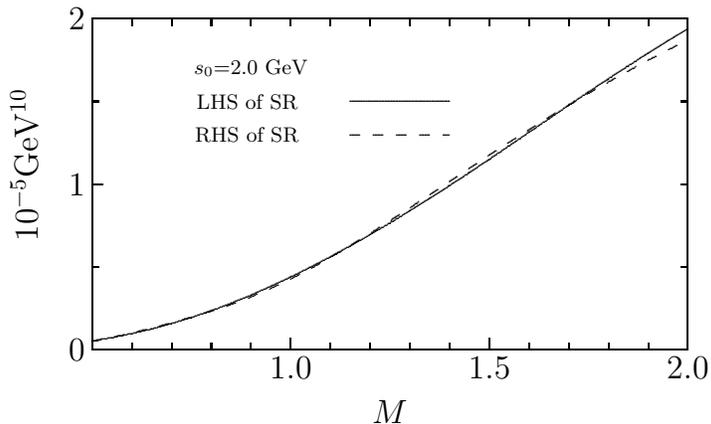,width=10cm,angle=0}}
\caption{The left hand side (solid line) and the right hand side
(dashed line) of the QCD sum rule with masses and residues
presented in Tab. 1 at $s_0=2.0$ GeV.} \label{fit}
\end{figure}

The Fig.\ref{ins}a  corresponds  to our single instanton current
which we discussed in Sec.II. This current could be related to the
phase of instanton liquid  with spontaneous chiral symmetry
breaking \cite{shuryak98}. The Fig.\ref{ins}b corresponds to the
chirality symmetric component of the QCD vacuum. Within the
instanton model one can consider it as contribution coming from
instanton--antiinstanton molecules. The latter component is
expected to give a small contribution at zero temperature but
might be important at temperatures above the deconfinement
temperature. Therefore, we will choose $\alpha=-\beta$ as a
``good" interpolating current for $\sigma$--tetraquark based on
the instanton picture of the QCD vacuum and the chirality
arguments given above \footnote{We should point out, that if we
will choose $\alpha=\beta$, the analysis of SR gives for the value
of the tetraquark  mass approximately the same as for the case of
``good" current for the threshold $s_0=2$~GeV. The physical
meaning of the small splitting of the tetraquark masses with two
completely different currents is difficult to explain for the
present.}. Assigning $\alpha=-\beta=1, \gamma=1/4$ to the total
current based on the structure of the instanton Lagrangian
Eq.(\ref{ratio}), we have stability in the sum rule thanks to the
cancelation in the contributions from the higher dimensional
operators and some part of the instanton effects. With the best
fit in the two resonance approximation, the fitted masses and the
residues are summarized in Tab.~1 and the good quality of the fit
is shown in Fig.~\ref{fit}. We interpret the lower mass in our SR
as the mass of $\sigma$-tetraquark and the higher mass as the mass
of its first radial excitation. According to the PDG \cite{PDG},
the candidates for the excited states are $f_0(1370), f_0(1500)$,
and $f_0(1710)$. We can see that the lower mass $m_1$ almost does
not depend on the value of the threshold and the higher mass $m_2$
looks similar to $f_0(1710)$. Indeed, since the quality of the fit
of the masses and the residues is best at $s_0=2.0$ GeV, we can
interpret the lower mass $\sim 780$ MeV as the mass of the
$\sigma$ and the higher mass $\sim 1775$ MeV as the mass of the
state $f_0(1710)$\footnote{Recently, this meson was reported to
have a mass $f_0(1790)$ by BES collaboration~\cite{BES}.}. In this
scheme the $f_0(1370)$ might be treated as a conventional
$q\bar{q}$ state and the $f_0(1500)$--meson can be consider as a
glueball candidate \cite{Anisovich05}.

\begin{table}[h]
\begin{center}
\begin{tabular}{|c|c|c|c|c|} \hline
$s_0$ (GeV) &  $m_{1}$ (GeV) &  $m_{2}$ (GeV) & $f_1 (10^{-3}$GeV) &
$f_2(10^{-3}$GeV)\\
\hline\hline
2.0 & 0.7822 & 1.7756 & 4.399 & 0.395\\
\hline
2.2 & 0.7964 & 1.9488 & 4.241 & 0.426\\
\hline
2.4 & 0.8102 & 2.1016 & 4.095 & 0.459\\
\hline
2.6 & 0.8261 & 2.2519 & 3.935 & 0.491\\
\hline
\end{tabular}
\caption{Fitted masses and residues in the two resonances
approximation with $\alpha=1, \beta=-1, \gamma=1/4$.}
\end{center}
\end{table}

\section{Conclusion}

We have discussed a novel interpolating current for scalar meson
$\sigma$(600) treated as a tetraquark state. The
color--spin--flavor structure of our current is fixed by the the
properties of the instanton induced quark--quark interaction and
reflects the topological structure of the QCD vacuum.
Our ``good" current has very peculiar chirality structure
because it includes both type of diquarks, scalar and pseudoscalar,
with equal weights. In this connection we would like
to point out that the similar improvement of the OPE convergence
for some specific currents was previously observed in the case of the
usual hadrons. Well known example is so--called Ioffe's current for
the nucleon which is widely using for description of the nucleon
properties \cite{ioffe2}. This current has also very peculiar
chirality structure and includes scalar and pseudoscalar diquarks
with equal weights.

We have demonstrated that the definite chirality structure of
our ``good" current for $\sigma$
leads to the cancelation of the high dimensional operator
contributions to OPE and to the vanishing of some instanton
contributions which can spoil the QCD sum rules.
As a result, within QCD sum rule approach, we have obtained a
very stable result for the
mass of the $\sigma$ meson around $780$ MeV. This mass lies inside
the range of the PDG $m_{\sigma(600)}=400\div 1200$ MeV. We should
also mention the possible change in the result for the mass if the
mixing between the usual two quark states and our tetraquark state
is taken into account. An additional shift of the $\sigma$--mass
might come from the large perturbative QCD corrections in
multiquark hadrons \cite{pivovarov}.

\section*{Acknowledgments}
We are happy to acknowledge useful discussions on various aspects
of this research with  A.P. Bakulev, A.E. Dorokhov, B.L. Ioffe,
A.V. Mikhailov, A.G. Oganesian and V. Vento. We are especially
grateful to Prof. D.-P. Min and Prof. S. Kim for their kind
hospitality at the School of Physics and Astronomy of Seoul
National University (NK) and Asia Pacific Center for Theoretical
Physics in the final stages of this work (NK, HJL). NK's work was
supported in part by the Brain Pool program of the Korea Research
Foundation through KOFST grant 042T--1--1, and by the Russian
Foundation for Basic Research through grant RFBR--04--02--16445.

\end{document}